\documentclass[]{emulateapj}

\usepackage{graphicx} 
    \usepackage{epstopdf} 
    \usepackage{longtable} 
    \usepackage{natbib}
    \usepackage{chngpage}
\begin{document}

\newcommand{\msun}{M$_{\sun}$}
\newcommand{\reff}{$R\rm{_{eff}}$}
\newcommand{\pc}{pseudocontinuum}
\newcommand{\asn}{SNR}
\newcommand{\asnm}{$\rm{\asn_{meas}}$}
\newcommand{\asnt}{$\rm{\asn_{target}}$}
\newcommand{\met}{[Z]}
\newcommand{\jband}{J\textendash{}band}
\newcommand{\ob}{h and $\chi$ Persei}
\newcommand{\obs}{Perseus OB-1}
\newcommand{\csq}{$\chi^2$}
\newcommand{\csqd}{$\chi^2_{{\rm data}}$}
\newcommand{\teff}{T$_{\rm{eff}}$}
\newcommand{\lte}{LTE}
\newcommand{\nlte}{NLTE}
\newcommand{\smalla}{a}

\newcommand{\hii}{H\,{\sc ii}\rm}
\newcommand{\oiii}{O\,{\sc iii}\rm}

\newcommand{\central}{$-0.03 \pm 0.05$}
\newcommand{\gradient}{$-0.44 \pm 0.08$}
\newcommand{\gradkpc}{$-0.083 \pm 0.014$}

\def\vrad{$v_{\rm rad}$}
\def\vrot{$v_{\rm rot}$}
\def\vpred{$v_{\rm pred}$}
\def\PA{{\it PA}}
\def\kms{km\,s$^{-1}$}
\def\deg{$^{\circ}$}
\def\teff{$T_{\rm eff}$}
\def\logg{$\log g$}

\title{Red Supergiants as Cosmic Abundance Probes: The Sculptor Galaxy NGC 300\footnotemark[\smalla]}

\author{J. Zachary Gazak\altaffilmark{1}, 
Rolf Kudritzki\altaffilmark{1,2},
Chris Evans\altaffilmark{3},
Lee Patrick\altaffilmark{4},
Ben Davies\altaffilmark{5}, 
Maria Bergemann\altaffilmark{6},
Bertrand Plez\altaffilmark{7},
Fabio Bresolin\altaffilmark{1},
Ralf Bender\altaffilmark{2},
Michael Wegner\altaffilmark{2},
Alceste Z. Bonanos\altaffilmark{8},
Stephen J. Williams\altaffilmark{8}}

\bibliographystyle{apj}  

\begin{abstract}

We present a quantitative spectroscopic study of twenty-seven red supergiants in the Sculptor Galaxy NGC 300.   \jband\ spectra were obtained using KMOS on the VLT and studied with state of the art synthetic spectra including \nlte\ corrections for the strongest diagnostic lines.  We report a central metallicity of \met = \central\ with a gradient of \gradkpc\ [dex/kpc$^{-1}$], in agreement with previous studies of blue supergiants and \hii-region auroral line measurements.  This result marks the first application of the \jband\ spectroscopic method to a population of individual red supergiant stars beyond the Local Group of galaxies and reveals the great potential of this technique.

\end{abstract}


\footnotetext[a]{Data collected under ESO Program ID 092.B-0088}

\altaffiltext{1}{Institute for Astronomy, University of Hawai'i, 2680 Woodlawn Dr, Honolulu, HI 96822, USA}
\altaffiltext{2}{University Observatory Munich, Scheinerstr. 1, D-81679 Munich, Germany}
\altaffiltext{3}{UK Astronomy Technology Centre, Royal Observatory Edinburgh, Blackford Hill, Edinburgh., EH9 3HJ, UK}
\altaffiltext{4}{Institute for Astronomy, Royal Observatory Edinburgh, Blackford Hill, Edinburgh., EH9 3HJ, UK}
\altaffiltext{5}{Astrophysics Research Institute, Liverpool John Moores University, 146 Brownlow Hill, Liverpool L3 5RF, UK}
\altaffiltext{6}{Institute of Astronomy, University of Cambridge, Madingley Road, Cambridge CB3 0HA, UK}
\altaffiltext{7}{Laboratoire Univers et Particules de Montpellier, Universit\'e de Montpellier, CNRS, F-34095 Montpellier, France}
\altaffiltext{8}{IAASARS, National Observatory of Athens, GR-15236 Penteli, Greece}

\maketitle

\section{Introduction}

The chemical enrichment of galaxies with heavy elements lays a complex roadway across the age of the star-forming universe, 
and techniques to measure metallicity are the astronomer's compass and sextant as the evolution of galaxies 
is mapped.  The primary tools of the trade so far have been the studies of the strongest emission lines of \hii\ regions. They
have provided important first hints at the enrichment history of the 
observable universe which include gradients across the disks of galaxies, a relationship between galactic 
mass and central metallicity, and metallicity evolution as a function of redshift.  Still, these strong-line \hii\ 
methods are plagued with systematics arising from their empirical calibrations and the complexity of the 
systems producing the efficiently observable emission lines used 
\citep{2008ApJ...681.1183K,2008ApJ...681..269K,2009ApJ...700..309B}.  At the same time, it
is a basic fact that the stellar population of galaxies provides the drive for chemical enrichment as well as the radiated light used to measure that 
process. Thus, the natural markers of metallicity are then the stars themselves.  

Only recently have developments in the modeling of stellar atmospheres, observational techniques, and 
statistical methods added the quantitative spectroscopy of supergiant stars to the extragalactic astronomer's toolset.  This 
pioneering work provided a measurement of the radial abundance gradient of NGC 300 using optical spectroscopy 
of blue supergiant stars and grids of synthetic spectra calculated without assuming local thermodynamic 
equilibrium (\lte).  This non-LTE (\nlte) quantitative technique revealed a central metallicity of slightly below 
solar and a clear gradient across the star-forming disk of the galaxy  \citep{2008ApJ...681..269K}.

Since that early work, blue supergiants have become a powerful tool for measuring metallicities, 
gradients, and distances to galaxies in and beyond the Local Group  
(WLM \textendash{} \citealt{2006ApJ...648.1007B,2008ApJ...684..118U}; 
NGC 3109 \textendash{} \citealt{2007ApJ...659.1198E,2014ApJ...785..151H}; 
IC1613 \textendash{} \citealt{2007ApJ...671.2028B}; 
M33 \textendash{} \citealt{2009ApJ...704.1120U}; 
M81 \textendash{} \citealt{2012ApJ...747...15K}; 
NGC 4258 \textendash{} \citealt{2013ApJ...779L..20K}; 
NGC 3621 \textendash{} \citealt{2014ApJ...788...56K}).  
As the technique continues to mature it shows good agreement with the 
chemical abundances obtained from the nebular emission line method
based on the determination of the gas temperature via the weak auroral lines 
\citep{2009ApJ...700..309B}, albeit most recent work  \citep{2013ApJ...779L..20K,2014ApJ...788...56K} 
indicates a small systematic difference of $\sim$0.1 dex.  
By comparison, metallicities derived from any of the host of strong line \hii\ methods 
return, at best, consistent trends with huge offsets in overall enrichment level.  

A growing body of observational work demonstrates that the quantitative spectroscopy of red supergiant (RSG) stars rivals the precision
of metallicity measurements using blue supergiants and is applicable over similar distance scales using existing
telescopes and instruments.  The pilot near-infrared \jband\ study of \cite{2010MNRAS.407.1203D} and high spectral resolution followup by
\cite{2014ApJ...788...58G} $-$both targeting galactic RSGs$-$demonstrate the applicability of the technique to stars with 
roughly solar chemical enrichment.  In \cite{rsg_lmcsmc_placeholder} and \cite{patricksub}, the technique is successfully applied to RSGs significantly below solar chemical abundance in the Small and Large Magellanic Clouds and the metal poor Local Group dwarf galaxy NGC 6822.  

These studies are aided by continued development of strongly improved quantitative synthetic spectra used to extract stellar
parameters.  While the underlying grid of {\sc marcs} stellar atmospheres is calculated in \lte\ \citep{2008A&A...486..951G}, corrections have
been developed for \nlte\ line formation of the strongest diagnostic lines (iron and titanium $-$ \cite{2012ApJ...751..156B}, silicon $-$ \cite{2013ApJ...764..115B},
and magnesium $-$ \cite{2014arXiv1412.6527B}).  

The RSG technique will become more important with the next generation telescopes such as the TMT and E$-$ELT.  These telescopes and their
instruments will be optimized for observations at infrared wavelengths, using adaptive optics supported multi object spectrographs. In consequence, investigating stars
radiating strongly in the near-IR$-$including red giant, asymptotic giant branch, and red supergiant stars $-$will have a clear advantage
 in the future.
 
In this paper we report another significant step. We have used the multi-IFU NIR spectrograph KMOS at the ESO VLT to test
the \jband\ RSG technique over a wide range of chemical enrichment across the star forming disk
of NGC 300 at a distance of 1.88 Mpc \citep{2005ApJ...628..695G}.  

This first application of the technique beyond the Local Group of galaxies is yet another milestone towards a large
scale application of the RSG technique to the understanding of the present day chemical enrichment of galaxies undergoing star formation.

After discussing the KMOS observations in \S\ref{sec:obs} we outline our analysis procedure in \S\ref{sec:technique}.  For a full description of the \jband\ technique we direct readers to \cite{2014ApJ...788...58G}.  Results and discussion are presented in \S\ref{sec:results}.

\begin{figure*}[tbp]
	\centering
	\includegraphics[width=17cm]{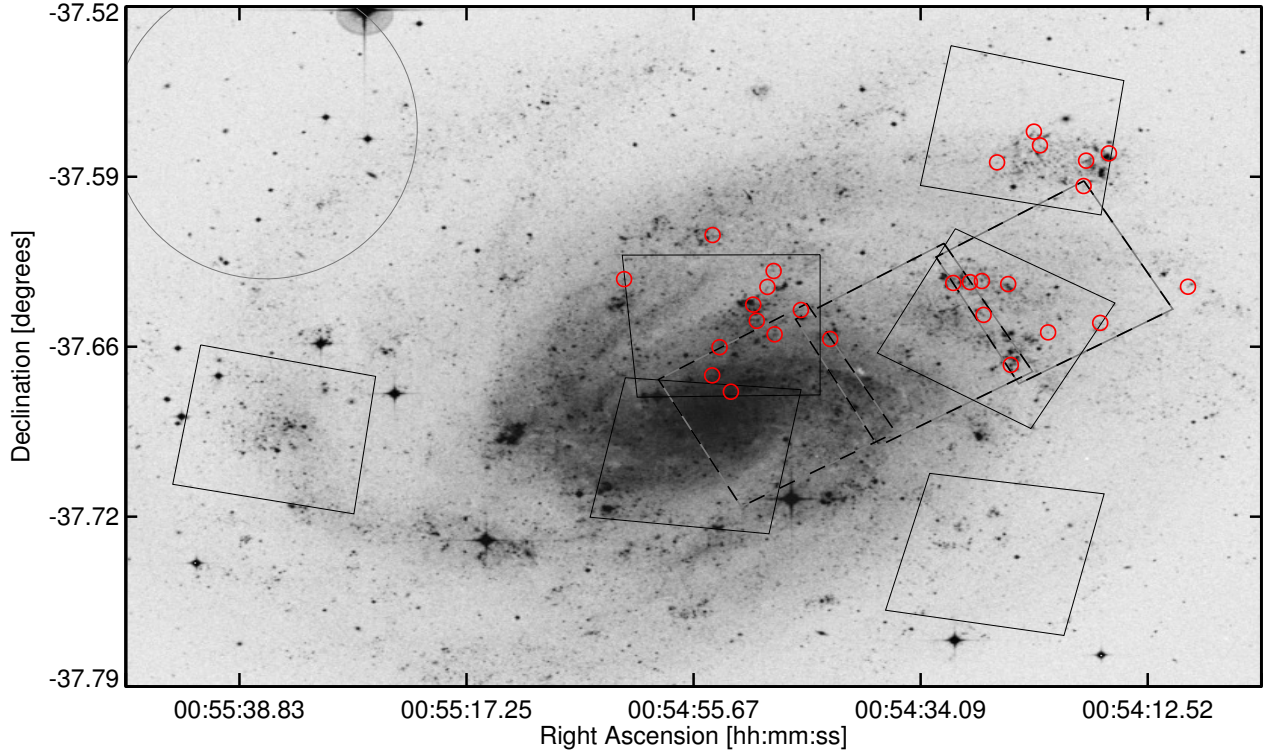}
	\caption{The spiral galaxy NGC300.  The Hubble ACS fields are overplotted in black, with WFPC2 frames in 
black dashed with gray.  Observed RSGs are circled in red.  In the upper left a gray circle denotes the size 
of the KMOS field of view, within which 24 IFUs can be placed per pointing.}
	\label{fig:n300_finder}
\end{figure*}

\begin{figure}[tbp]
	\centering
	\includegraphics[width=8.5cm]{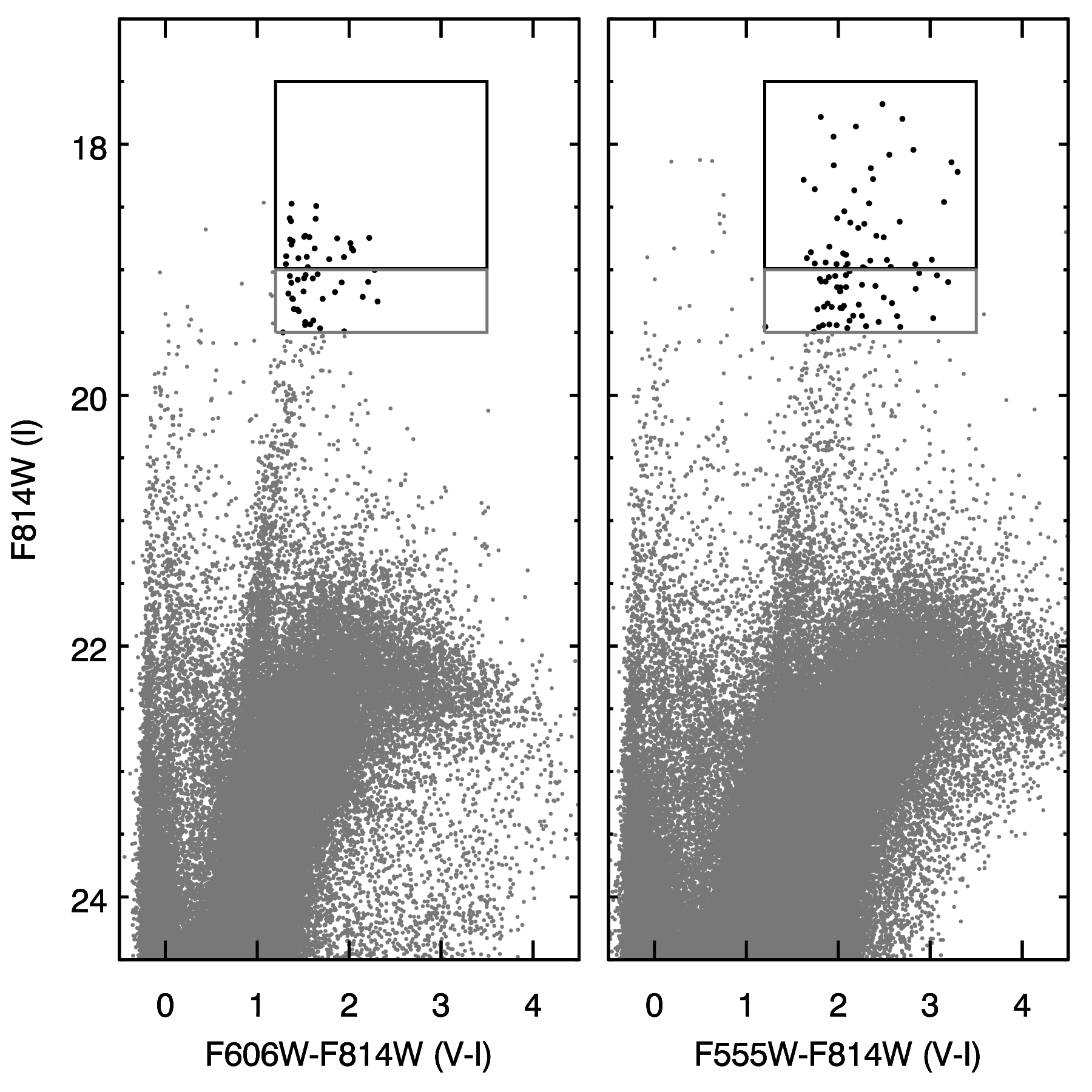}
	\caption{Color Magnitude selection for extragalactic RSGs.  The left panel shows color vs magnitude for the WFPC observations of NGC 300 while the right panel 
	is the same but for ACS fields.  The boxed regions show the brighter ``priority one" RSG candidates in this
	color magnitude space and the fainter ``priority two" region below.}
	\label{fig:n300_cmd}
\end{figure}

\section{Observations}
\label{sec:obs}
\subsection{Target Selection}

We prepared an initial database of RSG candidates using the ACS Nearby Galaxy 
Survey Treasury (ANGST: \citealt{2009ApJS..183...67D}), a public database of stellar 
photometry obtained with Hubble Space Telescope observations.  The ANGST 
catalog contains 6 fields in NGC 300 from the Advanced Camera for Surveys (ACS) 
and 3 fields observed with the Wide Field and Planetary Camera 2 (WFPC2). The layout 
of these fields can be seen in Figure~\ref{fig:n300_finder}.  We 
select the red F814W (``I'') filter for magnitude cuts.  The color cut for both instruments 
was ``V$-$I'', using F555W and F814W filters for ACS, F606W and F814W filters for 
WFPC2.  The cool temperatures and extreme luminosities of RSGs allow for high 
fidelity selection using such color-magnitude cuts.  This is demonstrated in 
Figure~\ref{fig:n300_cmd}: the ``RSG plume" (boxed) separates from the population 
of fainter red objects in m$_{I}$ magnitude dimension and is distinct from other bright, 
hotter objects with a red V$-$I color.  
Using the ANGST photometry we define RSG candidates as those objects with colors
bounded in V$-$I color between 1.2 and 3.5. and having m$_{I}$ brighter than 19.5, a 
value selected to achieve our target signal to noise ratio (SNR) of $\sim$100 \citep{2014ApJ...788...58G}.
Seeking the best spectra possible we divided these targets into two lists separated at m$_{I}$ $\geq$ 19.0 
and gave higher priority to the brighter group. 

We note that candidate objects within the fainter group could be, in principle, Asymptotic Giant Branch 
stars (AGB). Stellar evolution models predict some overlap between RGB and so-called super-AGB at 
log L/L$_\odot$ $\sim$4.5. However, these objects would also be young with ages smaller than 100 Myr 
with abundances of Fe, Si, Ti unaltered from their initial composition so that they would also be good 
tracers of the host galaxyÕs present  day abundances.

Because the HST coverage of NGC 300 is incomplete we used overlaps in our 
database with a more complete but shallower catalog of B and V photometry \citep{2001A&A...371..497P}
to train a B$-$V vs m$_{V}$ color-magnitude cut.  Candidates selected by this
method were placed in a third ranked priority group such that HST selected targets would 
be observed first but that no IFUs would go unused.  

As a final adjustment to our RSG candidate list we search for overlaps between our existing
candidates and Spitzer photometry of NGC 300 from \cite{2010ApJ...715.1094K}.  The important
Spitzer diagnostic for RSG candidacy is the color-magnitude plane defined using IRAC Band 1
(m$_{3.6}$) and the color Band 1 $-$ Band 2 ([3.6]$-$[4.5]).  RSGs separate as the brightest 
m$_{3.6}$ targets with color blue ward of 0.0 in [3.6]$-$[4.5] \citep{2009A&A...498..127V,2009AJ....138.1003B}.  
We upgrade all objects in our candidate list with overlap as Spitzer RSG candidates to the top priority group.  Coordinates and photometric data of our targets finally used for the analysis are provided in Table~\ref{tbl:fitpar}.

\subsection{Observations}

The resulting candidate catalog was used as input to the KMOS ARM Allocator (KARMA) 
software for planning the setup of the 24 KMOS Integral Field Units (IFUs) \citep{2008SPIE.7019E..0TW}. These 2$\arcsec$.8 
square IFUs can be placed over the 7$\arcmin$.2 KMOS field of view (shown in 
Figure~\ref{fig:n300_finder}).  We observed an inner and outer field to cover a significant range 
of the radial extent of NGC300.  These fields were observed in 600s integrations over the nights of 
2013 October 14, 15, and 16.  The 14th and 15th nights were clear and stable at Paranal with 
median seeing of $\approx$1$\arcsec$.1.  Weather degraded slightly on the night of the 16th with 
intermittent cloud cover and variable seeing with the median value around 1$\arcsec$.3.

KMOS was operated in nod to sky mode with science integrations of 600 seconds.  Telluric standards 
were observed down each IFU at a frequency of once per every 60 to 90 minutes.  KMOS IFU data 
cubes were flat fielded, wavelength calibrated, and telluric corrected using the KMOS 
pipeline (v1.3.2) provided publicly as a specific instance of the European Southern Observatory (ESO) 
Reflex automated data reduction environment \citep{2013A&A...559A..96F}.  

\begin{figure}
\begin{centering}
\includegraphics[width=9cm]{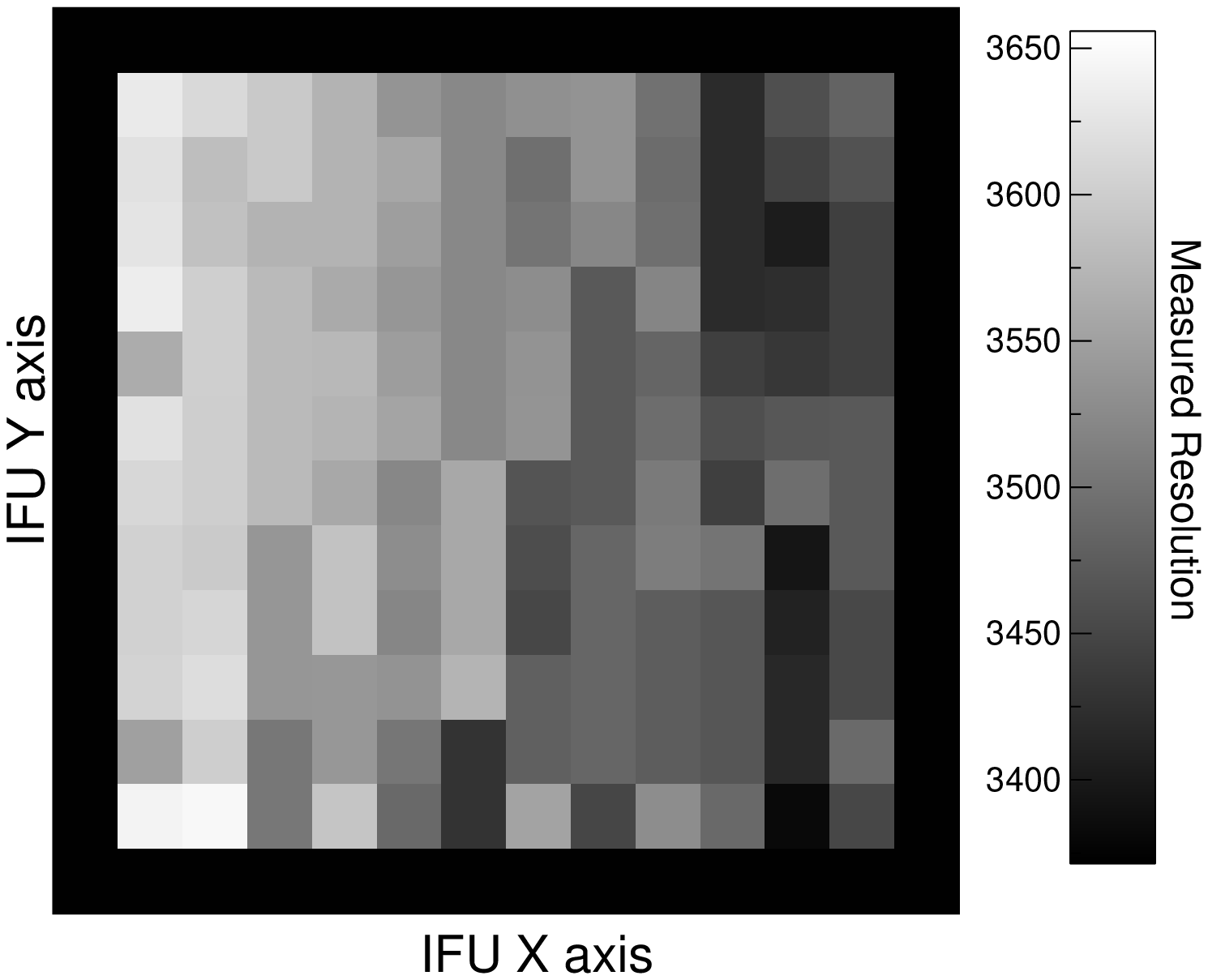}\\
\includegraphics[width=9cm]{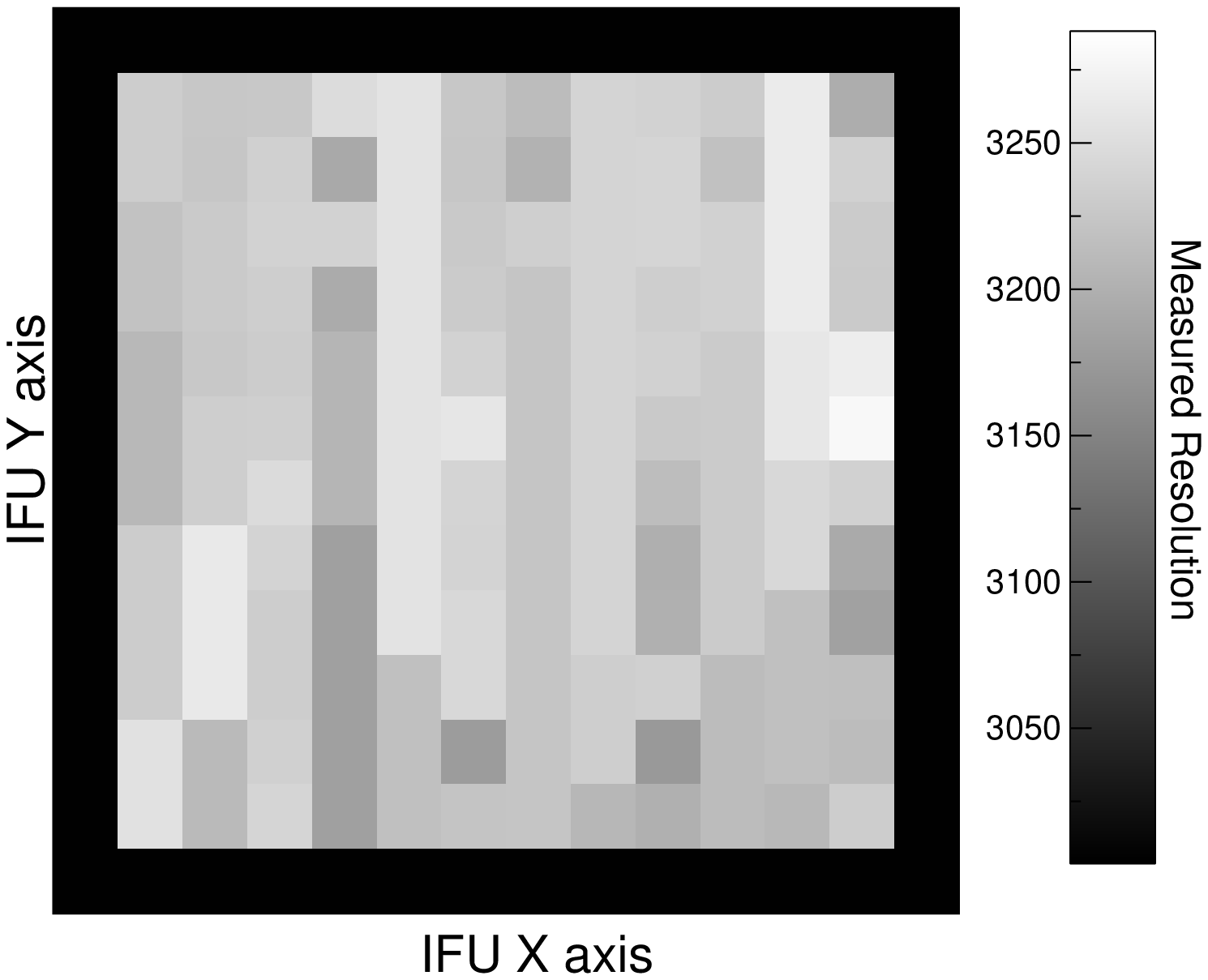}
\caption{Variation of spectral resolution of the inner 12x12 spatial pixels of a typical KMOS IFU measured from night sky lines. The top panel
shows a typical measurement after a 600 sec exposure. The bottom panel presents the resolution after the homogenizing procedure described in the text is applied.}
\label{fig:skyfit}
\end{centering}
\end{figure}

We find inhomogeneities in spectral resolution across the field of view of each KMOS IFU as shown 
in Figure~\ref{fig:skyfit}.  To correct for this we first measure the effective spectral resolution of night
sky lines in each spatial pixel of the inner 12 x 12 IFU field.  Each sky line is well approximated
by a gaussian and for each spatial pixel we calculate the weighted mean of the full width half maxima
of those gaussian models.  The top panel of Figure~\ref{fig:skyfit} shows one such example.  The 
effective resolutions, in this case, vary by up to $\Delta$R$\sim$300.  We correct this by smoothing the sky and
science spectra for each IFU to bring the field to a more constant resolution of $\Delta$R$\sim$100 lower than the minimum
measured effective resolution.  The results of re-measuring effective resolution are displayed in the 
bottom panel of Figure~\ref{fig:skyfit}, which has a global offset in measured resolution but is plotted
on the same dynamic scale.  The effective resolution after these corrections ranges from the expected value after smoothing to values smaller by
$\Delta$R$\sim$100.

Once the data cubes are homogenized we extract a science spectrum.  We do this by summing over each wavelength
slice in the IFU data cube.  In practice, we find that the highest signal to noise is recovered by extracting in a two pixel 
radius around the peak target flux in the science frame.  We suspect that inhomogeneities in wavelength across the 
field of view play some role in this.    

After constructing 1D spectra of object and sky frames, we utilize the ESO tool Skycorr which
takes as input those two spectra and scales airglow lines using theoretical knowledge of the
night sky lines.  The resulting sky spectrum provides a removal of variability in sky emission based on difference
in time and sky position of the two observations \citep{2014A&A...567A..25N}.

\begin{figure*}[tbp]
	\begin{centering}
	\includegraphics[width=15cm]{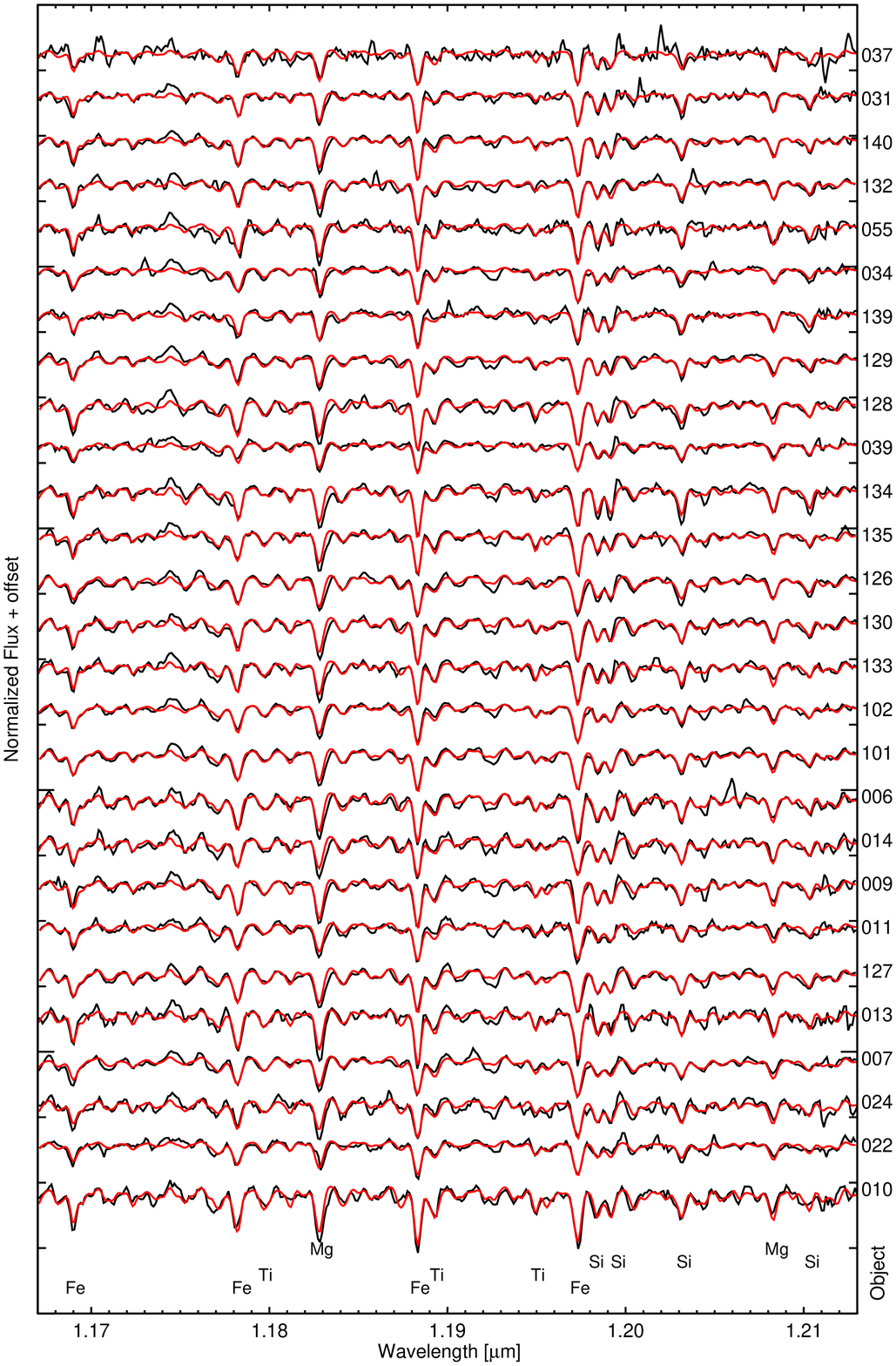}
	\caption{NGC 300 RSG spectra plotted in black with corresponding best model fits in red.  Diagnostic features are
	marked.  Each object is labeled to the right of the plotted axis and information corresponding to the fit is tabulated by this name in
	Table~\ref{tbl:results}.}
	\label{fig:atlas1}
	\end{centering}
\end{figure*}

As each field was observed multiple times, the final science spectra were produced  by extracting the median 
of all individual sky-subtracted spectra.  This spectral atlas is plotted in Figure~\ref{fig:atlas1}.

\subsection{Target validation}

\begin{figure}
\includegraphics[width=8.5cm]{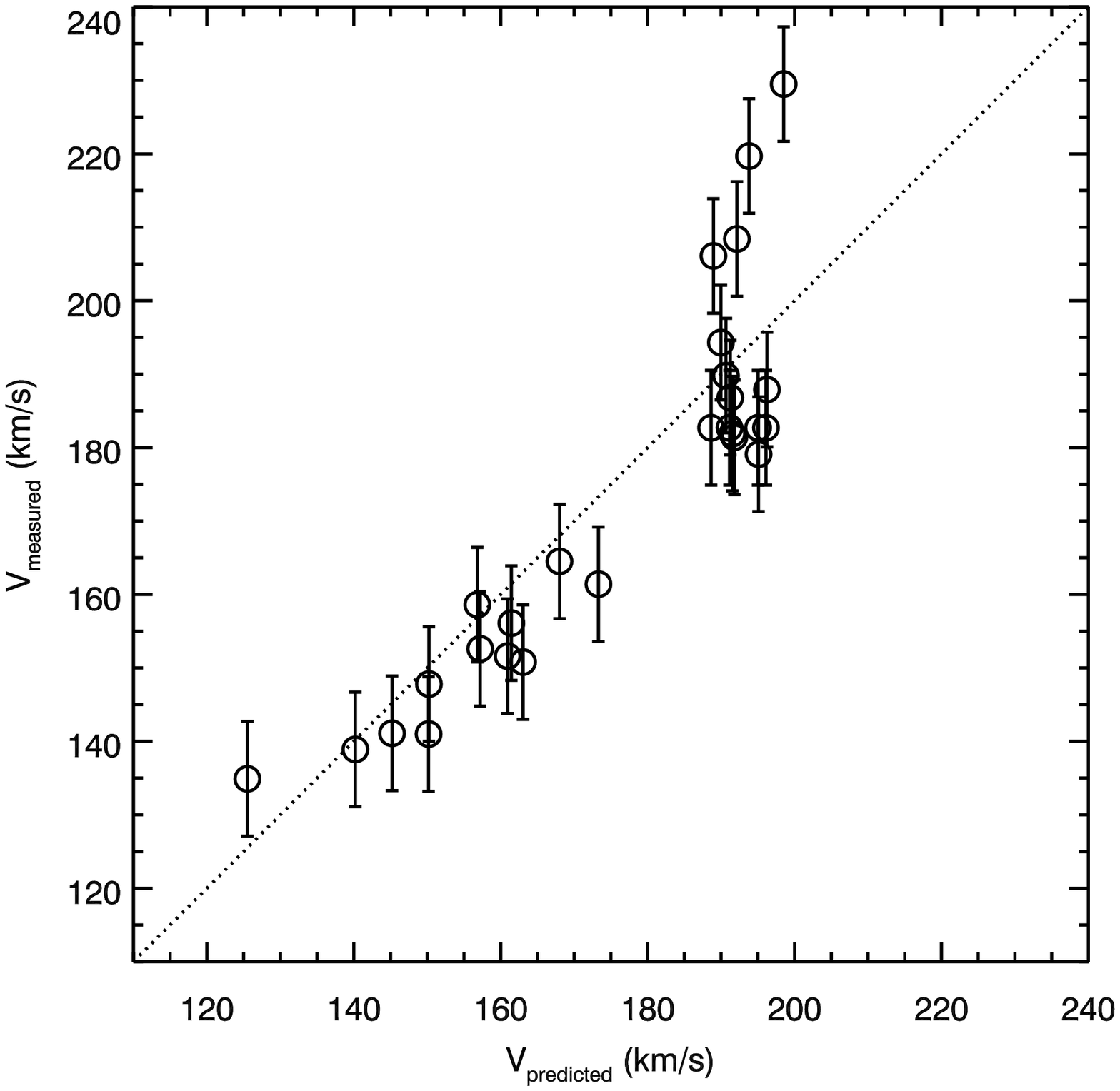}
\caption{The observed radial velocities of our sample stars versus
  that predicted using the model rotation curve of NGC~300 from \citet{2011MNRAS.410.2217W}.}
\label{vrfig}
\end{figure}

In order to verify that our selected targets were indeed members of
NGC~300 and not foreground objects, we studied the radial velocities \vrad\
of each of the stars. We measured \vrad\ by cross-correlating the
reduced spectra with a model spectrum from our grid. The choice of
model spectrum was found to be unimportant, producing variations in
the measured \vrad\ of less than 1\kms, insignificant compared to the
other sources of error (see below). The \vrad\ were then corrected for
barycentric motion and converted to the Local Standard of Rest ({\it
LSR}) using the Starlink package {\sc rv}. 

For the errors on \vrad\ we first specified a lower limit to the
uncertainty of 1/10th of a pixel, which at the 2-pixel sampling of
KMOS corresponds to $\pm$5\kms. To this error we added in quadrature
the observed spatial variations in wavelength calibration across each
IFU, which we found to be $\pm$6\kms. This gives us an error on each
\vrad\ of $\pm$8\kms.

To determine what the {\it expected} \vrad\ would be at the position
of each target, we used the H\,{\sc i} map of \citet{2011MNRAS.410.2217W}, and
their model of NGC~300's rotation curve. These authors fit the
observed velocity map with a model which allowed the rotational
velocity \vrot, inclination $i$ and position angle \PA\ to vary with
galactocentric distance. Since the variation in $i$ was found to be
small (45$\pm$5\deg), we kept this parameter fixed at 45\deg. For
\vrot\ (=40-100\kms) and \PA (=290-330\deg), we used fits to their
observed trends with galactocentric distance. The predicted radial
velocity \vpred\ as a function of position on the plane of the sky
relative to the centre of the galaxy ($x,y$) was then determined from,

\begin{equation}
v_{\rm pred}(x,y) = v_{\rm sys} + v_{\rm rot} \cos \theta \sin i
\end{equation}

\noindent where $v_{\rm sys}=136$\kms\ \citep{2011MNRAS.410.2217W}, and
$\cos\theta$ is given by,

\begin{equation}
\cos\theta = \frac{ -x \sin({\it PA}) + y\cos({\it PA})}{r}
\end{equation}

\indent and,

\begin{equation}
\displaystyle
r = \left[  x^2 + \left(\frac{y}{\cos i}\right)^2 \right]
\end{equation}

In Fig.\ \ref{vrfig} we plot the observed versus predicted radial
velocities for each star in our sample. The first thing we can say is
that there are no obvious candidates for foreground stars in our
sample, since such objects would be expected to have low radial
velocities of a few $\times$10\kms. For the majority of our sample,
there is excellent agreement between their measured radial velocities
and the model rotation curve to within 1$\sigma$. The only deviations
greater than 2$\sigma$ are seen at high \vrad, where four datapoints
lie $\sim$20\kms\ above the predicted trend. If NGC~300 is similar to
the Milky Way in that deviations from the model rotation curve of
20\kms\ are not uncommon \citep{2003A&A...397..133R}, then we can say all stars
in our sample have radial velocities consistent with being NGC~300
members.

\begin{figure*}[tbp]
\begin{centering}
\includegraphics[width=16cm]{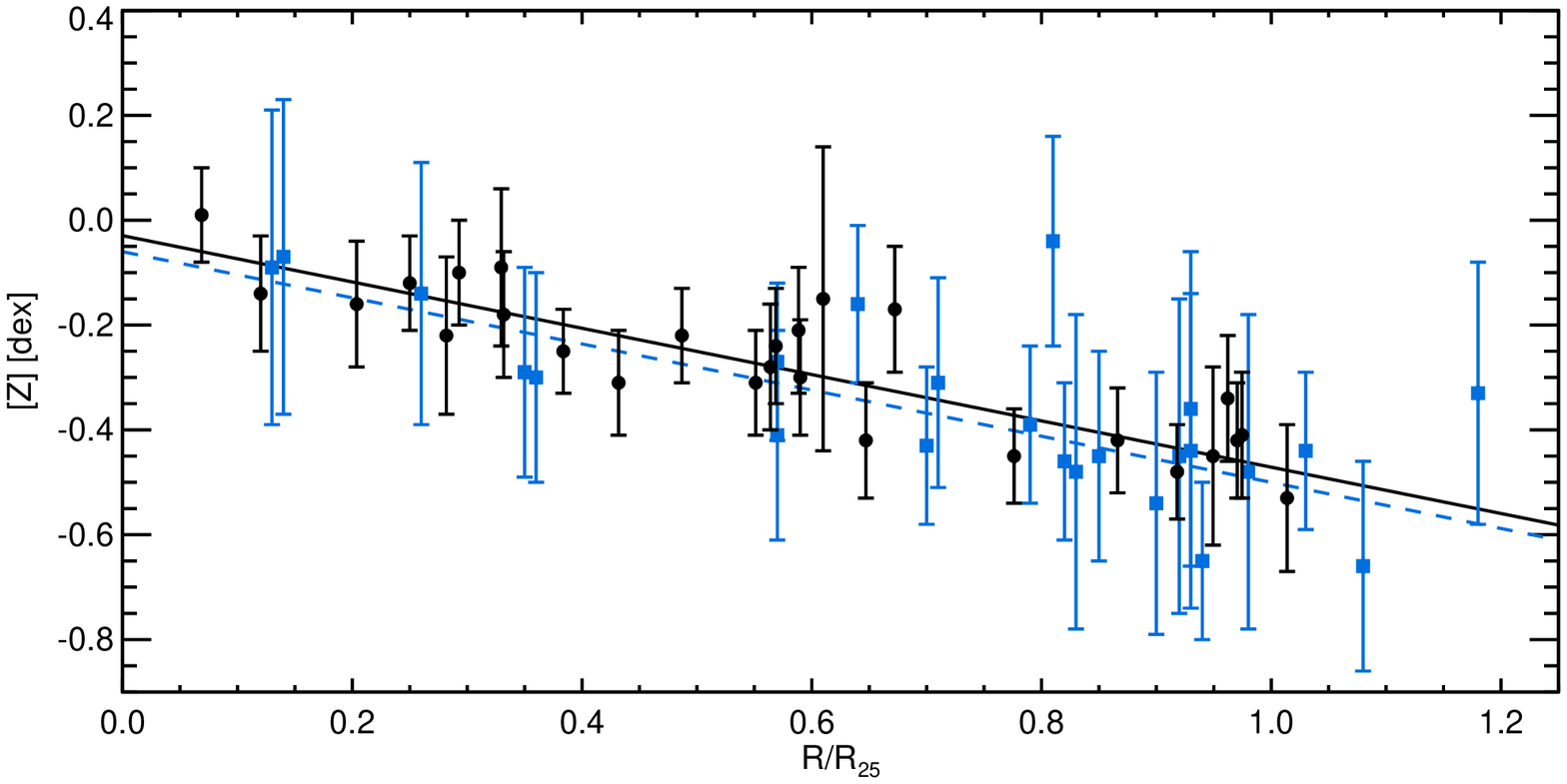}
\caption{The stellar metallicities and metallicity gradients in NGC 300 at 1.88 Mpc.  Blue squares are BSGs from  \cite{2008ApJ...681..269K} and a linear regression through those points is plotted as a blue dashed line.  Black circles and the corresponding solid black linear regression line are RSGs from this work.}
\label{fig:n300grad}
\end{centering}
\end{figure*}

\section{Analysis}
\label{sec:technique}

Observed spectra are analyzed through a comparison with synthetic spectra based on a grid of 1D \lte\  {\sc marcs} 
model atmospheres \citep{2008A&A...486..951G}.  The line formation calculations of the synthetic spectra are carried 
out in \nlte\ for the elements iron, titanium, silicon and magnesium, which produce the most critical diagnostic lines
prominent in the J-band spectra. The details of the \nlte\ radiative transfer method and atomic models are described 
in  \cite{2012ApJ...751..156B,2013ApJ...764..115B,2014arXiv1412.6527B}. All other lines including the (weak) molecular 
contributions are calculated in \lte.  The free parameters of this grid
include effective temperature, log gravity, metallicity [Z] (normalized to Solar values, [Z] = log Z/Z$_{\odot}$), and microturbulence 
(\teff, log$g$, \met, $\xi$ $-$ see Table~$\ref{tbl:par_grid}$ for parameter coverage).
For the analysis we apply the same technique as described in detail in \cite{2014ApJ...788...58G}. We note, however, that we use an updated set of synthetic spectra
calculated with improved oscillator strength for the J-band TiI lines (see \citealt{rsg_lmcsmc_placeholder}).

For a discussion of the limitations and possible improvements of the model atmospheres and the radiative transfer 
methods used we refer to the publications cited above and the references therein. On the other hand, after the 
successful tests of the method at Milky Way \citep{2014ApJ...788...58G} and significantly lower metallicities 
\citep{rsg_lmcsmc_placeholder,patricksub} we are confident that the method delivers reliable metallicities.

\begin{deluxetable}{lcccc}
\centering
\tabletypesize{\small}
\tablewidth{0pt}
\tablecaption{{\sc marcs} Model Grid}
\tablehead{
\colhead{Parameter}  & \colhead{Notation} & \colhead{Min}   & \colhead{Max}  & \colhead{Spacing}}
\startdata
Eff. Temperature  [K]  & \teff  & 3400  & 4000  & 100 \\
 &  &  4000  & 4400 & 200 \\
Log gravity & log$g$ & $-$1.0  & +1.0  &  0.5 \\
Metallicity [dex] & \met\  & $-$1.00 & +1.00  & 0.25 \\
Microturbulence [km/s] & $\xi$ & 1.0  & 6.0   &  1.0 
\enddata
\tablecomments{Parameter grid for {\sc marcs} atmospheres (and synthetic spectra) utilized in this work.}
\label{tbl:par_grid}
\end{deluxetable}

In addition to model parameters we fit for ``effective resolution'', \reff, an unresolvable combination of 
the effects of instrument spectral resolution R ($\delta\lambda / \lambda$) of each IFU and a very small contribution of macro turbulence in the 
atmospheres of RSGs.   This is done by degrading each model to a set of resolutions ranging from twice 
the expected spectral resolution downwards until a clear \csq\ minimum has been defined.  The minimum of 
a parabolic fit through the \csq\ vs. R is adopted as the best fit \reff\ for that combination of model and
data.  After an initial construction of the \csq\ grid allowing each model to be measured with its own ideal
 \reff\ we lock the  \reff\ to the value which matches the minimum of that \csq\ and re-calculate.  This step 
 is completed for accurate measurement of parameter uncertainties.

Correcting the continuum level of our models to that of the observed spectra starts with selecting the highest model
flux point in wavelength bins of characteristic ``continuum width'' $CW = \lambda / 2R_{eff}$ such that the 
fit samples the full spectrum.  We note that at the low resolution of spectra in this work, the true
continuum is lost to a blended forest of molecular lines such that we are actually fitting a ``\pc''.  While the 
level of the \pc\ is certainly a function of the physical stellar parameters, especially \met, we find that no
detrimental effects due to this phenomenon appear to resolutions significantly below that of our observed
spectra (see \citealt{2014ApJ...788...58G}).  After removing continuum outliers we fit a smooth third order polynomial through
the ratio of observed flux to model flux at the continuum points as a function of wavelength.  This function is
then applied to scale model to data.

With the continuum normalization procedure described in the preceding paragraph it is important to realize 
that our spectroscopic \jband\ method is not affected by the potential presence of circumstellar dust. Though 
such dust may add a small amount of extinction, it will affect the pseudo continuum and the lines equally and, 
thus, leave the abundance measurements unaltered.   

We measure best fit parameters by isolating the six 2D planes in each parameter pair combination which contain
the minimum \csq.  As a result, two parameters are locked to the best fit values in each of the six slices.  We
interpolate the \csq\ grid of each plane to a parameter grid with four times the density and take the minimum
of that dense grid as the best fit values for the two parameters defining the plane.  After completing the procedure 
for all six planes, the three measurements of each parameter are averaged into a set of best fit parameters 
(again we refer to \citealt{2014ApJ...788...58G} for all details). 

Parameter uncertainties are assessed with a Monte Carlo simulation.  This is accomplished by adding 1000
instances of random gaussian noise to a linear interpolation from the model grid to the best fit parameters.  
The gaussian noise is scaled such that the SNR of this noisy model matches the SNR of the observed 
spectrum.  For each instance we determine fit parameters and for each parameter we define a 1$\sigma$ 
uncertainty such that the central 68\% of the 1000 measured parameters lies within $\pm 1\sigma$.  
The best fit parameters and their derived uncertainties are presented in Table~\ref{tbl:results}.

\section{Results and Discussion}
\label{sec:results}

\begin{deluxetable*}{cllrlrr}
\tablewidth{0pt}
\tabletypesize{\scriptsize}
\tablecaption{\nlte\ Stellar Parameters for NGC 300 Red Supergiants}
\tablehead{
\colhead{Target}
& \colhead{R/R$_{25}$}
& \colhead{T$_{\rm{eff}}$ [K]}
& \colhead{log$g$}
& \colhead{Z [dex]}
& \colhead{$\xi$ [km/s]}
& \colhead{R [$\frac{\lambda}{\delta \lambda}$]}}
\startdata
  010 & 0.069 & 4096 $\pm$  40 & $-$0.88 $\pm$ 0.2 & $+$0.01 $\pm$ 0.09 & 4.4 $\pm$ 0.4 & 3500 \\
  022 & 0.120 & 3535 $\pm$  70 & $-$0.73 $\pm$ 0.3 & $-$0.14 $\pm$ 0.11 & 3.1 $\pm$ 0.2 & 2800 \\
  024 & 0.204 & 4153 $\pm$  40 & $-$0.16 $\pm$ 0.2 & $-$0.16 $\pm$ 0.12 & 4.0 $\pm$ 0.2 & 3000 \\
  007 & 0.250 & 3912 $\pm$  70 & $-$0.52 $\pm$ 0.2 & $-$0.12 $\pm$ 0.09 & 4.0 $\pm$ 0.2 & 2700 \\
  013 & 0.282 & 3875 $\pm$  50 & $-$0.47 $\pm$ 0.5 & $-$0.22 $\pm$ 0.15 & 4.3 $\pm$ 0.2 & 3500 \\
  127 & 0.293 & 4220 $\pm$  30 & $-$0.80 $\pm$ 0.3 & $-$0.10 $\pm$ 0.10 & 4.3 $\pm$ 0.3 & 2800 \\
  011 & 0.330 & 4063 $\pm$  70 & $+$0.47 $\pm$ 0.2 & $-$0.09 $\pm$ 0.15 & 4.1 $\pm$ 0.2 & 2800 \\
  009 & 0.332 & 4075 $\pm$  40 & $-$0.12 $\pm$ 0.2 & $-$0.18 $\pm$ 0.12 & 4.3 $\pm$ 0.2 & 3100 \\
  014 & 0.384 & 4260 $\pm$  30 & $-$0.38 $\pm$ 0.2 & $-$0.25 $\pm$ 0.08 & 3.6 $\pm$ 0.2 & 3400 \\
  006 & 0.432 & 4205 $\pm$  90 & $-$0.42 $\pm$ 0.3 & $-$0.31 $\pm$ 0.10 & 4.1 $\pm$ 0.2 & 3200 \\
  101 & 0.487 & 4228 $\pm$  42 & $-$0.52 $\pm$ 0.1 & $-$0.22 $\pm$ 0.09 & 4.2 $\pm$ 0.2 & 2800 \\
  102 & 0.551 & 4010 $\pm$ 147 & $+$0.07 $\pm$ 0.2 & $-$0.31 $\pm$ 0.10 & 3.9 $\pm$ 0.2 & 2800 \\
  133 & 0.564 & 3953 $\pm$  52 & $-$0.40 $\pm$ 0.1 & $-$0.28 $\pm$ 0.12 & 3.6 $\pm$ 0.2 & 3500 \\
  130 & 0.569 & 4123 $\pm$  30 & $-$0.44 $\pm$ 0.2 & $-$0.24 $\pm$ 0.11 & 4.1 $\pm$ 0.2 & 3000 \\
  126 & 0.589 & 4251 $\pm$  60 & $-$0.41 $\pm$ 0.2 & $-$0.21 $\pm$ 0.12 & 4.1 $\pm$ 0.2 & 2900 \\
  135 & 0.590 & 3953 $\pm$  53 & $-$0.32 $\pm$ 0.2 & $-$0.30 $\pm$ 0.11 & 3.8 $\pm$ 0.3 & 3200 \\
  134 & 0.610 & 4400 $\pm$  20 & $-$0.17 $\pm$ 0.7 & $-$0.15 $\pm$ 0.29 & 5.0 $\pm$ 0.2 & 3400 \\
  039 & 0.647 & 4240 $\pm$  50 & $+$0.56 $\pm$ 0.3 & $-$0.42 $\pm$ 0.11 & 3.0 $\pm$ 0.2 & 3000 \\
  128 & 0.672 & 3776 $\pm$  63 & $-$0.53 $\pm$ 0.4 & $-$0.17 $\pm$ 0.12 & 5.0 $\pm$ 0.2 & 3000 \\
  129 & 0.776 & 4113 $\pm$  50 & $-$0.28 $\pm$ 0.2 & $-$0.45 $\pm$ 0.09 & 4.1 $\pm$ 0.3 & 2900 \\
  139 & 0.866 & 4280 $\pm$  30 & $+$0.56 $\pm$ 0.4 & $-$0.42 $\pm$ 0.10 & 4.1 $\pm$ 0.2 & 3100 \\
  034 & 0.918 & 3833 $\pm$  44 & $-$0.01 $\pm$ 0.2 & $-$0.48 $\pm$ 0.09 & 4.1 $\pm$ 0.2 & 2500 \\
  055 & 0.949 & 3981 $\pm$ 119 & $-$0.40 $\pm$ 0.1 & $-$0.45 $\pm$ 0.17 & 4.1 $\pm$ 0.2 & 3100 \\
  132 & 0.962 & 4160 $\pm$  43 & $-$0.26 $\pm$ 0.2 & $-$0.34 $\pm$ 0.12 & 4.5 $\pm$ 0.2 & 2800 \\
  140 & 0.970 & 4141 $\pm$  30 & $+$0.04 $\pm$ 0.3 & $-$0.42 $\pm$ 0.11 & 4.1 $\pm$ 0.2 & 3200 \\
  031 & 0.975 & 4270 $\pm$  40 & $+$0.20 $\pm$ 0.2 & $-$0.41 $\pm$ 0.12 & 4.0 $\pm$ 0.2 & 3400 \\
  037 & 1.014 & 3912 $\pm$  50 & $+$0.40 $\pm$ 0.2 & $-$0.53 $\pm$ 0.14 & 3.1 $\pm$ 0.3 & 2900
\enddata
\tablecomments{R$_{25}$ = 5.33 kpc (Bresolin et al., 2009) based on a distance of 1.88 Mpc \citep{2005ApJ...628..695G}.}
\label{tbl:results}
\end{deluxetable*}

\subsection{Results}

The parameter fits are in general quite precise with individual metallicities as accurate as 0.10 to 0.15 dex. The exception is object 134 with a high effective
temperature just at the edge of our grid of synthetic spectra. It may well be that this object is slightly hotter which would reduce its metallicity and improve the fit of the silicon lines.

In  the following, we discuss metallicity and metallicity gradient and the evolutionary status of our objects. The main results are summarized in Figures~\ref{fig:n300grad},~\ref{fig:n300teffmet},~\ref{fig:n300hrd}.

\subsection{Metallicity Gradient}

This first measurement of metallicity of the young stellar population of stars across the disk of a spiral galaxy beyond the Local Group using the RSG J-band method requires a 
careful comparison with previous work.

The most direct comparison of our metallicity results is to the work by \cite{2008ApJ...681..269K}, who studied 24 A- and B-type 
supergiants across the radial extent of this galaxy.   They used synthetic \nlte\ spectra to derive 
metallicities from an ensemble of heavy atomic species (Mg, Si, S, Ti, Cr, and Fe) and derived logarithmic metallicities [Z] relative to the Sun. We display their results
in Figure~\ref{fig:n300grad}. With a linear regression we obtain a central metallicity of $-$0.07 $\pm$ 0.09 dex and a gradient of $-0.081 \pm 0.011$ dex kpc$^{-1}$ from this sample of BSGs.  
We note that the original \cite{2008ApJ...681..269K} values for central metallicity and gradient were $-$0.06 dex and 
$-$0.083 dex kpc$^{-1}$, respectively.  We applied the new galactic orientation model of \cite{2009ApJ...700..309B} to
calculate slightly different galactocentric distances.  This resulted in a small change in the central metallicity and metallicity gradient.

Our \jband\ method is a natural extension of the  \cite{2008ApJ...681..269K} technique and the results for our observed RSG sample are also shown in Figure~\ref{fig:n300grad}. We 
stress that we use an independent but
evolutionarily connected population of stars, as blue supergiants below masses of $\sim$30\msun\ evolve into the 
RSGs we observe as they exhaust their core hydrogen.  By all means, then, the metallicities obtained with the two techniques should agree.  Still,
the stellar models and synthetic spectral calculations are independent for our  two techniques as the physical conditions
of these two populations are distinct.  The cool, inflated atmospheres of RSGs contain a host of singly ionized metals
and a forest of molecular features which are absent in the hot spectra of BSGs.  We also observe our stars in a different
wavelength regime subject to separate observational difficulties.  Finally, due to the added complexity of RSG
atmospheres it is not yet possible to perform calculations in pure \nlte, even though our grid of synthetic spectra do have
\nlte\ calculations for the most critical spectral features.  Despite these differences, the \jband\
technique derives metallicity as a ratio of the solar abundance pattern just as the BSG method does, using strong atomic lines of Fe, Ti, Si, 
and Mg.  Our observations of NGC300 RSGs then become an excellent test of both methods and of the theoretical 
calculations from which they draw. 

Using our observations and the RSG \jband\ technique, we measure a central metallicity of \central\ dex and a gradient of \gradkpc\ dex kpc$^{-1}$.  This is a stunning agreement
between BSGs and the RSG \jband\ method and certainly represents a breakthrough of the technique.  

\cite{2009ApJ...700..309B} provide an important second measurement of the central metallicity and
gradient of NGC 300 using measurements of \hii\ region auroral lines.  Studies using this ``direct method'' of 
measurement agree with abundances derived from populations of stars in environments where metallicity 
is below solar, for example, dwarf galaxies \citep{2005A&A...434..507S,2006ApJ...648.1007B,2006ApJ...642..813L}, and for some regions in the Milky Way and M33 (MW: \citealt{2000A&A...363..537R,2000MNRAS.311..329D,0004-637X-617-2-1115}, M33:  \citealt{1988MNRAS.235..633V,0004-637X-635-1-311,2011ApJ...730..129B}). 
In this technique the [\oiii]
auroral lines give access to a key physical parameter, the electron temperature $T_e$, which can disentangle the
effect of line strengths based on oxygen abundance and temperature.  \cite{2009ApJ...700..309B} find a central
oxygen abundance of 12+log(O/H) = 8.57 $\pm$ 0.02 and a gradient of $-$0.077 $\pm$ 0.006 dex kpc$^{-1}$.  The
gradient is slightly shallower than for the RSGs and BSGs, but the difference is small.  The comparison of central
metallicity depends on the assumed value for the solar oxygen abundance.  Choosing 12+log(O/H)$_\odot$ = 8.69  
\citep{2009ARA&A..47..481A} returns a \hii\ region central metallicity value of Z = $-$0.12 $\pm$ 0.02.  While this
value agrees with the BSG result it indicates a small offset ($\sim$0.09 dex) relative to the RSGs.  We note, however,
that such an offset is also found in the recent work by \cite{2013ApJ...779L..20K,2014ApJ...788...56K} in the 
spiral galaxies NGC 4258 and 3621 when BSG and $T_e$ \hii\ region metallicities are compared. A possible reason for this discrepancy is the depletion of nebular oxygen on dust grains as, for instance, suggested by \cite{2012MNRAS.427.1463Z}. On the other hand, we note that comparing the metallicities obtained with the three methods at R/R$_{25}$ = 1  the situation looks much better. We find [Z] = $-$0.50, $-$0.47, $-$0.53 for the BSG, RSG, and \hii\ auroral line methods, respectively. Given that the three methods are entirely different the agreement is striking.  The measurements from this work are tabulated with the discussed work from the literature in Table~\ref{tbl:workvlit}.

\subsection{Stellar Evolution}

\begin{figure}[tbp]
	\begin{centering}
	\includegraphics[width=8.5cm]{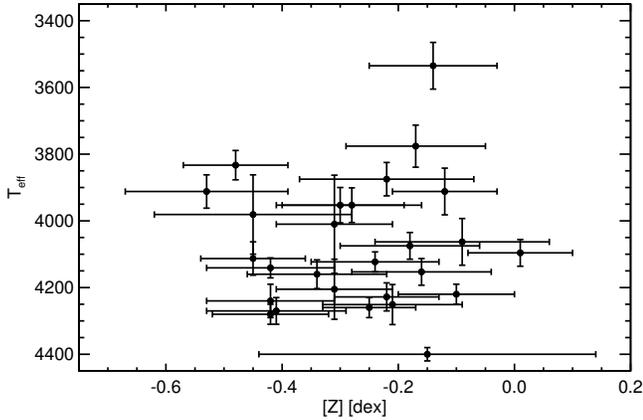}
	\caption{Effective temperature of our RSG targets as a function of metallicity.}
	\label{fig:n300teffmet}
	\end{centering}
\end{figure}

Stellar evolution theory predicts a shift of the Hayashi limit (the minimum temperature of an RSG evolutionary track, when stars are nearly fully convective) towards higher effective
temperature with decreasing metallicity. For a stellar evolutionary track with solar metallicity and a mass of M = 15 M$_{\odot}$ a minimum temperature of $\sim$3600K is obtained \citep{2012A&A...537A.146E,2002A&A...390..561M},
whereas models with SMC like metallicity 
\citep{2013A&A...558A.103G}
have a minimum temperature of $\sim$4100K. In contrast to this prediction, the RSG J-band spectroscopy in the Milky Way, LMC \& SMC, and NGC 6822 
\citep{2014ApJ...788...58G,rsg_lmcsmc_placeholder,patricksub}
does not show such a trend and leads to the conclusion that the temperatures of RSGs are independent of metallicity.
We can use our observations in NGC 300 as an additional test in this regard since we encounter a range of metallicities between solar in the center of the NGC 300 disk down to 1/3 solar
at the isophotal radius. Accordingly, Figure~\ref{fig:n300teffmet} shows a plot of effective temperature against metallicity. It confirms the absence of a trend at least in the metallicity
range covered.  In 1-D evolutionary models, the temperature of the Hayashi limit
(i.e. the maximum size of a star) is governed by the treatment of
convection, in particular by the mixing-length parameter $\alpha$. The
value of $\alpha$ is typically tuned in order to reproduce the
properties of the Sun, but then is fixed at this value for stars of
all masses, ages and metallicities. This fixed value of $\alpha$
typically results in evolutionary paths for RSGs which are `inclined',
decreasing in \teff\ as $L$ increases, as well as producing trends of
lower \teff\ for RSGs with higher initial masses, and higher average
\teff\ at lower metallicities \citep{2000A&A...361..101M,2008MNRAS.384.1109E,2011A&A...530A.115B}.

Recently, studies using 3-D hydrodynamical simulations of convection
have attempted to empirically model the evolution of $\alpha$ for
stars across a range of \teff, \logg\ and metallicity
\citep{2014MNRAS.445.4366T,2015A&A...573A..89M}. The nature of the simulations used in
these studies meant that it was not possible to cover the parameter
space occupied by RSGs, and instead focussed on stars of higher
gravities and temperatures. However, extrapolating from the observed
trends in \citet{2015A&A...573A..89M}, the results seem to suggest that $\alpha$
should increase at lower metallicity for a fixed \teff\ and \logg. If
correct, this would cause RSGs at a fixed initial mass and
evolutionary stage to become {\it hotter} at lower metallicity,
exacerbating the trend seen in evolutionary models rather than
counteracting it. In effect this is the opposite to what we find in
our results, that the average effective temperature of RSGs seems to
be independent of metallicity. 

The discussion of Figure~\ref{fig:n300teffmet} is, of course, based on the assumption that 
our determination of effective temperatures using the \jband\ technique is not affected by 
spurious systematic effects which lead to erroneous effective temperatures which 
are always very similar even for significantly different metallicities. A first clear indication 
that this is not the case comes from the work by \cite{2013ApJ...767....3D}. This comprehensive 
study of RSGs in the LMC and SMC determined effective temperatures using ESO VLT 
XShooter observations of SEDs covering the full spectral range from the near-UV to the near-IR. 
It was found that the RSG effective temperatures in both galaxies are very similar and do not 
seem to be affected by the difference in metallicity. Moreover, \cite{rsg_lmcsmc_placeholder} 
recently re-investigated the same sample of RSGs applying the spectroscopic \jband\ technique 
and found effective temperatures consistent with the fits of full SEDs indicating that the \jband\ 
method provides robust estimates of effective temperatures. For a more detailed discussion of 
RSG temperature determinations we refer to \cite{2013ApJ...767....3D} and \cite{2014ApJ...788...58G}.

\begin{figure}[tbp]
	\begin{centering}
	\includegraphics[width=8.5cm]{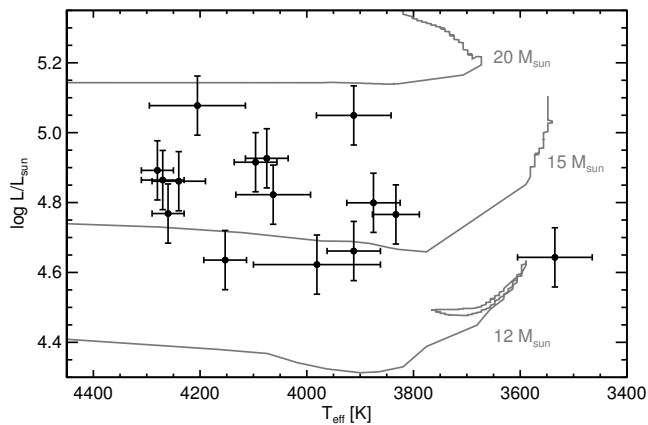}
	\caption{Hertzsprung-Russell diagram of program stars for which we have F814W (I) band photometry from Hubble.  Temperatures
	are the parameter fits from our analysis procedure and we calculate luminosities using the bolometric correction recipes for RSGs 
	of \cite{2013ApJ...767....3D}.}
	\label{fig:n300hrd}
	\end{centering}
\end{figure}

We construct a Hertzsprung-Russell diagram (HRD) to compare our results with stellar evolution theory.
We calculate bolometric luminosities for program stars using F814W (I) band photometry from Hubble and
the bolometric correction recipes of \cite{2013ApJ...767....3D}.  We adopt the distance modulus of $\mu$ = 
26.37 $\pm$ 0.05 from the work by \cite{2008ApJ...681..269K} and \cite{2005ApJ...628..695G}.  These luminosities are 
plotted against the effective temperatures from our 
spectral fit in the HRD of Figure~\ref{fig:n300hrd}. We then overplot 
evolutionary tracks with solar metallicity adopting the Geneva database of 
stellar evolutionary models including the effects of 
rotation \citep{2000A&A...361..101M}.  We note that all program stars with HST magnitudes fall well within the 
ranges in temperature and luminosity which are appropriate for RSGs.  We calculate stellar masses by interpolating 
fitted temperatures and gravities and calculated luminosities to the Geneva evolutionary models.  Errors in mass are 
calculated with a monte carlo simulation in which, for 1000 trials, we add noise to \teff, \logg, and luminosity on the 
scale of our fit uncertainties.  The mass values and errors presented in Table $\ref{tbl:fitpar}$ represent the median and
1$\sigma$ standard deviation of those monte carlo experiments. This experiment indicates an initial mass range between 12 and 20 M$_{\odot}$.

\begin{deluxetable*}{lccccl}
\tablewidth{0pt}
\tablecaption{Comparison of NGC 300 Metallicity and Gradient Literature}
\tablehead{
\colhead{Study}
& \multicolumn{2}{c}{Central Abundance} 
& \multicolumn{2}{c}{Metallicity Gradient} 
& \colhead{Notes}
\\ 
& dex, metals
& 12+log(O/H)
& R/R$_{25}$
& dex kpc$^{-1}$
& }
\startdata
\cite{2008ApJ...681..269K} &  $-0.07 \pm 0.09$  &  \nodata  & $-0.44 \pm 0.06$   & $-0.081 \pm 0.011$    &   Blue Supergiants, Metals \\
\cite{2009ApJ...700..309B} & \nodata                         &  8.57 $\pm$ 0.02  & $-0.41 \pm 0.03$  & $-0.077 \pm 0.006$ & \hii\ regions, auroral oxygen \\ 
This work                                & \central                         &                                 & \gradient                      & \gradkpc                           & Red Supergiants, Metals 
\enddata
\tablecomments{Three independent measurements of the evolution of chemical abundance across the star forming disk of NGC 300.  The agreement
between BSGs, RSGs, and auroral \hii\ region methods is excellent.}
\label{tbl:workvlit}
\end{deluxetable*}

\section{Conclusions and Future Work}

The quantitative spectroscopy of individual stars in distant galaxies 
as a way to constrain the evolution of galaxies has long been a dream 
of stellar astronomers. We believe that with the
results presented here we have made a major step forward to turn 
this dream into reality. The stunning agreement between two completely 
independent stellar spectroscopic methods 
analyzing massive stars in different evolutionary phases with  
different sets of model atmospheres, spectral ranges, ionization 
stages and spectral lines clearly indicates that 
extragalactic stellar spectroscopy has matured to become an 
accurate and powerful tool. With new multi-object NIR spectrographs 
such as KMOS at the VLT and MOSFIRE at Keck and the ``classical''
MOS instruments at visual light for BSGs such as FORS/VLT and 
LRIS/Keck we have now two reliable tools to investigate the chemical 
evolution of galaxies  out to $\sim$7 Mpc.
With future telescopes with 30 to 40m apertures as the TMT or 
E-ELT and adaptive optics (AO) supported NIR multi-object spectrographs 
the RSG J-band method can be pushed much further. \cite{2011A&A...527A..50E}
have demonstrated that for AO supported KMOS-like spectrographs magnitudes 
down to J$\sim$24 mag could be reached with sufficient signal-to-noise in 
two nights of observing. This will allow studies like the one presented here 
to be carried out at 50 Mpc distance and will make a significant volume of the local universe accessible for accurate 
quantitative investigations of the chemical evolution of galaxies.


\section{Acknowledgments}
JZG and RPK acknowledge support by the National Science Foundation under grant AST-1108906 and the hospitality of the Munich University Observatory where part of this work was carried out.  BD is supported by a fellowship from the Royal Astronomical Society. BP is supported in part by the Programme National de Physique Stellaire of the INSU CNRS. SJW and AZB acknowledge funding by the European Union (European Social Fund) and National Resources under the ÒARISTEIAÓ action of the Operational Programme ÒEducation and Lifelong LearningÓ in Greece.

\bibliography{massive_stars}

\begin{thebibliography}{53}
\expandafter\ifx\csname natexlab\endcsname\relax\def\natexlab#1{#1}\fi

\bibitem[{{Asplund} {et~al.}(2009){Asplund}, {Grevesse}, {Sauval}, \&
  {Scott}}]{2009ARA&A..47..481A}
{Asplund}, M., {Grevesse}, N., {Sauval}, A.~J., \& {Scott}, P. 2009, \araa, 47,
  481

\bibitem[{{Bergemann} {et~al.}(2014){Bergemann}, {Kudritzki}, {Gazak},
  {Davies}, \& {Plez}}]{2014arXiv1412.6527B}
{Bergemann}, M., {Kudritzki}, R.-P., {Gazak}, Z., {Davies}, B., \& {Plez}, B.
  2014, ArXiv e-prints

\bibitem[{{Bergemann} {et~al.}(2012){Bergemann}, {Kudritzki}, {Plez}, {Davies},
  {Lind}, \& {Gazak}}]{2012ApJ...751..156B}
{Bergemann}, M., {Kudritzki}, R.-P., {Plez}, B., {Davies}, B., {Lind}, K., \&
  {Gazak}, Z. 2012, \apj, 751, 156

\bibitem[{{Bergemann} {et~al.}(2013){Bergemann}, {Kudritzki}, {W{\"u}rl},
  {Plez}, {Davies}, \& {Gazak}}]{2013ApJ...764..115B}
{Bergemann}, M., {Kudritzki}, R.-P., {W{\"u}rl}, M., {Plez}, B., {Davies}, B.,
  \& {Gazak}, Z. 2013, \apj, 764, 115

\bibitem[{{Bonanos} {et~al.}(2009){Bonanos}, {Massa}, {Sewilo}, {Lennon},
  {Panagia}, {Smith}, {Meixner}, {Babler}, {Bracker}, {Meade}, {Gordon},
  {Hora}, {Indebetouw}, \& {Whitney}}]{2009AJ....138.1003B}
{Bonanos}, A.~Z., {et~al.} 2009, \aj, 138, 1003

\bibitem[{{Bresolin}(2011)}]{2011ApJ...730..129B}
{Bresolin}, F. 2011, \apj, 730, 129

\bibitem[{{Bresolin} {et~al.}(2009){Bresolin}, {Gieren}, {Kudritzki},
  {Pietrzy{\'n}ski}, {Urbaneja}, \& {Carraro}}]{2009ApJ...700..309B}
{Bresolin}, F., {Gieren}, W., {Kudritzki}, R., {Pietrzy{\'n}ski}, G.,
  {Urbaneja}, M.~A., \& {Carraro}, G. 2009, \apj, 700, 309

\bibitem[{{Bresolin} {et~al.}(2006){Bresolin}, {Pietrzy{\'n}ski}, {Urbaneja},
  {Gieren}, {Kudritzki}, \& {Venn}}]{2006ApJ...648.1007B}
{Bresolin}, F., {Pietrzy{\'n}ski}, G., {Urbaneja}, M.~A., {Gieren}, W.,
  {Kudritzki}, R., \& {Venn}, K.~A. 2006, \apj, 648, 1007

\bibitem[{{Bresolin} {et~al.}(2007){Bresolin}, {Urbaneja}, {Gieren},
  {Pietrzy{\'n}ski}, \& {Kudritzki}}]{2007ApJ...671.2028B}
{Bresolin}, F., {Urbaneja}, M.~A., {Gieren}, W., {Pietrzy{\'n}ski}, G., \&
  {Kudritzki}, R. 2007, \apj, 671, 2028

\bibitem[{{Brott} {et~al.}(2011){Brott}, {de Mink}, {Cantiello}, {Langer}, {de
  Koter}, {Evans}, {Hunter}, {Trundle}, \& {Vink}}]{2011A&A...530A.115B}
{Brott}, I., {et~al.} 2011, \aap, 530, A115

\bibitem[{Daflon \& Cunha(2004)}]{0004-637X-617-2-1115}
Daflon, S., \& Cunha, K. 2004, The Astrophysical Journal, 617, 1115

\bibitem[{{Dalcanton} {et~al.}(2009){Dalcanton}, {Williams}, {Seth}, {Dolphin},
  {Holtzman}, {Rosema}, {Skillman}, {Cole}, {Girardi}, {Gogarten},
  {Karachentsev}, {Olsen}, {Weisz}, {Christensen}, {Freeman}, {Gilbert},
  {Gallart}, {Harris}, {Hodge}, {de Jong}, {Karachentseva}, {Mateo}, {Stetson},
  {Tavarez}, {Zaritsky}, {Governato}, \& {Quinn}}]{2009ApJS..183...67D}
{Dalcanton}, J.~J., {et~al.} 2009, \apjs, 183, 67

\bibitem[{{Davies} {et~al.}(2010){Davies}, {Kudritzki}, \&
  {Figer}}]{2010MNRAS.407.1203D}
{Davies}, B., {Kudritzki}, R., \& {Figer}, D.~F. 2010, \mnras, 407, 1203

\bibitem[{{Davies} {et~al.}(2014){Davies}, {Kudritzki}, {Gazak}, {Plez},
  {Bergemann}, {Evans}, \& {Patrick}}]{rsg_lmcsmc_placeholder}
{Davies}, B., {Kudritzki}, R., {Gazak}, J.~Z., {Plez}, B., {Bergemann}, M.,
  {Evans}, C., \& {Patrick}, L. 2014, \apj

\bibitem[{{Davies} {et~al.}(2013){Davies}, {Kudritzki}, {Plez}, {Trager},
  {Lan{\c c}on}, {Gazak}, {Bergemann}, {Evans}, \&
  {Chiavassa}}]{2013ApJ...767....3D}
{Davies}, B., {et~al.} 2013, \apj, 767, 3

\bibitem[{{Deharveng} {et~al.}(2000){Deharveng}, {Pe{\~n}a}, {Caplan}, \&
  {Costero}}]{2000MNRAS.311..329D}
{Deharveng}, L., {Pe{\~n}a}, M., {Caplan}, J., \& {Costero}, R. 2000, \mnras,
  311, 329

\bibitem[{{Ekstr{\"o}m} {et~al.}(2012){Ekstr{\"o}m}, {Georgy}, {Eggenberger},
  {Meynet}, {Mowlavi}, {Wyttenbach}, {Granada}, {Decressin}, {Hirschi},
  {Frischknecht}, {Charbonnel}, \& {Maeder}}]{2012A&A...537A.146E}
{Ekstr{\"o}m}, S., {et~al.} 2012, \aap, 537, A146

\bibitem[{{Eldridge} {et~al.}(2008){Eldridge}, {Izzard}, \&
  {Tout}}]{2008MNRAS.384.1109E}
{Eldridge}, J.~J., {Izzard}, R.~G., \& {Tout}, C.~A. 2008, \mnras, 384, 1109

\bibitem[{{Evans} {et~al.}(2007){Evans}, {Bresolin}, {Urbaneja},
  {Pietrzy{\'n}ski}, {Gieren}, \& {Kudritzki}}]{2007ApJ...659.1198E}
{Evans}, C.~J., {Bresolin}, F., {Urbaneja}, M.~A., {Pietrzy{\'n}ski}, G.,
  {Gieren}, W., \& {Kudritzki}, R. 2007, \apj, 659, 1198

\bibitem[{{Evans} {et~al.}(2011){Evans}, {Davies}, {Kudritzki}, {Puech},
  {Yang}, {Cuby}, {Figer}, {Lehnert}, {Morris}, \&
  {Rousset}}]{2011A&A...527A..50E}
{Evans}, C.~J., {et~al.} 2011, \aap, 527, A50+

\bibitem[{{Freudling} {et~al.}(2013){Freudling}, {Romaniello}, {Bramich},
  {Ballester}, {Forchi}, {Garc{\'{\i}}a-Dabl{\'o}}, {Moehler}, \&
  {Neeser}}]{2013A&A...559A..96F}
{Freudling}, W., {Romaniello}, M., {Bramich}, D.~M., {Ballester}, P., {Forchi},
  V., {Garc{\'{\i}}a-Dabl{\'o}}, C.~E., {Moehler}, S., \& {Neeser}, M.~J. 2013,
  \aap, 559, A96

\bibitem[{{Gazak} {et~al.}(2014){Gazak}, {Davies}, {Kudritzki}, {Bergemann}, \&
  {Plez}}]{2014ApJ...788...58G}
{Gazak}, J.~Z., {Davies}, B., {Kudritzki}, R., {Bergemann}, M., \& {Plez}, B.
  2014, \apj, 788, 58

\bibitem[{{Georgy} {et~al.}(2013){Georgy}, {Ekstr{\"o}m}, {Eggenberger},
  {Meynet}, {Haemmerl{\'e}}, {Maeder}, {Granada}, {Groh}, {Hirschi}, {Mowlavi},
  {Yusof}, {Charbonnel}, {Decressin}, \& {Barblan}}]{2013A&A...558A.103G}
{Georgy}, C., {et~al.} 2013, \aap, 558, A103

\bibitem[{{Gieren} {et~al.}(2005){Gieren}, {Pietrzy{\'n}ski}, {Soszy{\'n}ski},
  {Bresolin}, {Kudritzki}, {Minniti}, \& {Storm}}]{2005ApJ...628..695G}
{Gieren}, W., {Pietrzy{\'n}ski}, G., {Soszy{\'n}ski}, I., {Bresolin}, F.,
  {Kudritzki}, R.-P., {Minniti}, D., \& {Storm}, J. 2005, \apj, 628, 695

\bibitem[{{Gustafsson} {et~al.}(2008){Gustafsson}, {Edvardsson}, {Eriksson},
  {J{\o}rgensen}, {Nordlund}, \& {Plez}}]{2008A&A...486..951G}
{Gustafsson}, B., {Edvardsson}, B., {Eriksson}, K., {J{\o}rgensen}, U.~G.,
  {Nordlund}, {\AA}., \& {Plez}, B. 2008, \aap, 486, 951

\bibitem[{{Hosek} {et~al.}(2014){Hosek}, {Kudritzki}, {Bresolin}, {Urbaneja},
  {Evans}, {Pietrzy{\'n}ski}, {Gieren}, {Przybilla}, \&
  {Carraro}}]{2014ApJ...785..151H}
{Hosek}, Jr., M.~W., {et~al.} 2014, \apj, 785, 151

\bibitem[{{Kewley} \& {Ellison}(2008)}]{2008ApJ...681.1183K}
{Kewley}, L.~J., \& {Ellison}, S.~L. 2008, \apj, 681, 1183

\bibitem[{{Khan} {et~al.}(2010){Khan}, {Stanek}, {Prieto}, {Kochanek},
  {Thompson}, \& {Beacom}}]{2010ApJ...715.1094K}
{Khan}, R., {Stanek}, K.~Z., {Prieto}, J.~L., {Kochanek}, C.~S., {Thompson},
  T.~A., \& {Beacom}, J.~F. 2010, \apj, 715, 1094

\bibitem[{{Kudritzki} {et~al.}(2008){Kudritzki}, {Urbaneja}, {Bresolin},
  {Przybilla}, {Gieren}, \& {Pietrzy{\'n}ski}}]{2008ApJ...681..269K}
{Kudritzki}, R., {Urbaneja}, M.~A., {Bresolin}, F., {Przybilla}, N., {Gieren},
  W., \& {Pietrzy{\'n}ski}, G. 2008, \apj, 681, 269

\bibitem[{{Kudritzki} {et~al.}(2014){Kudritzki}, {Urbaneja}, {Bresolin},
  {Hosek}, \& {Przybilla}}]{2014ApJ...788...56K}
{Kudritzki}, R.-P., {Urbaneja}, M.~A., {Bresolin}, F., {Hosek}, Jr., M.~W., \&
  {Przybilla}, N. 2014, \apj, 788, 56

\bibitem[{{Kudritzki} {et~al.}(2012){Kudritzki}, {Urbaneja}, {Gazak},
  {Bresolin}, {Przybilla}, {Gieren}, \&
  {Pietrzy{\'n}ski}}]{2012ApJ...747...15K}
{Kudritzki}, R.-P., {Urbaneja}, M.~A., {Gazak}, Z., {Bresolin}, F.,
  {Przybilla}, N., {Gieren}, W., \& {Pietrzy{\'n}ski}, G. 2012, \apj, 747, 15

\bibitem[{{Kudritzki} {et~al.}(2013){Kudritzki}, {Urbaneja}, {Gazak}, {Macri},
  {Hosek}, {Bresolin}, \& {Przybilla}}]{2013ApJ...779L..20K}
{Kudritzki}, R.-P., {Urbaneja}, M.~A., {Gazak}, Z., {Macri}, L., {Hosek}, Jr.,
  M.~W., {Bresolin}, F., \& {Przybilla}, N. 2013, \apjl, 779, L20

\bibitem[{{Lee} {et~al.}(2006){Lee}, {Skillman}, \&
  {Venn}}]{2006ApJ...642..813L}
{Lee}, H., {Skillman}, E.~D., \& {Venn}, K.~A. 2006, \apj, 642, 813

\bibitem[{{Magic} {et~al.}(2015){Magic}, {Weiss}, \&
  {Asplund}}]{2015A&A...573A..89M}
{Magic}, Z., {Weiss}, A., \& {Asplund}, M. 2015, \aap, 573, A89

\bibitem[{{Mennickent} {et~al.}(2004){Mennickent}, {Pietrzy{\'n}ski}, \&
  {Gieren}}]{2004MNRAS.350..679M}
{Mennickent}, R.~E., {Pietrzy{\'n}ski}, G., \& {Gieren}, W. 2004, \mnras, 350,
  679

\bibitem[{{Meynet} \& {Maeder}(2000)}]{2000A&A...361..101M}
{Meynet}, G., \& {Maeder}, A. 2000, \aap, 361, 101

\bibitem[{{Meynet} \& {Maeder}(2002)}]{2002A&A...390..561M}
---. 2002, \aap, 390, 561

\bibitem[{{Noll} {et~al.}(2014){Noll}, {Kausch}, {Kimeswenger}, {Barden},
  {Jones}, {Modigliani}, {Szyszka}, \& {Taylor}}]{2014A&A...567A..25N}
{Noll}, S., {Kausch}, W., {Kimeswenger}, S., {Barden}, M., {Jones}, A.~M.,
  {Modigliani}, A., {Szyszka}, C., \& {Taylor}, J. 2014, \aap, 567, A25

\bibitem[{{Patrick} {et~al.}(2014){Patrick}, {Evans}, {Davies}, {Kudritzki},
  {Gazak}, {Bergemann}, {Plez}, \& {Ferguson}}]{patricksub}
{Patrick}, L., {Evans}, C., {Davies}, B., {Kudritzki}, R., {Gazak}, J.~Z.,
  {Bergemann}, M., {Plez}, B., \& {Ferguson}, A. 2014, \apj, in press

\bibitem[{{Pietrzy{\'n}ski} {et~al.}(2001){Pietrzy{\'n}ski}, {Gieren},
  {Fouqu{\'e}}, \& {Pont}}]{2001A&A...371..497P}
{Pietrzy{\'n}ski}, G., {Gieren}, W., {Fouqu{\'e}}, P., \& {Pont}, F. 2001,
  \aap, 371, 497

\bibitem[{{Rolleston} {et~al.}(2000){Rolleston}, {Smartt}, {Dufton}, \&
  {Ryans}}]{2000A&A...363..537R}
{Rolleston}, W.~R.~J., {Smartt}, S.~J., {Dufton}, P.~L., \& {Ryans}, R.~S.~I.
  2000, \aap, 363, 537

\bibitem[{{Russeil}(2003)}]{2003A&A...397..133R}
{Russeil}, D. 2003, \aap, 397, 133

\bibitem[{{Skrutskie} {et~al.}(2006){Skrutskie}, {Cutri}, {Stiening},
  {Weinberg}, {Schneider}, {Carpenter}, {Beichman}, {Capps}, {Chester},
  {Elias}, {Huchra}, {Liebert}, {Lonsdale}, {Monet}, {Price}, {Seitzer},
  {Jarrett}, {Kirkpatrick}, {Gizis}, {Howard}, {Evans}, {Fowler}, {Fullmer},
  {Hurt}, {Light}, {Kopan}, {Marsh}, {McCallon}, {Tam}, {Van Dyk}, \&
  {Wheelock}}]{2006AJ....131.1163S}
{Skrutskie}, M.~F., {et~al.} 2006, \aj, 131, 1163

\bibitem[{{Stasi{\'n}ska}(2005)}]{2005A&A...434..507S}
{Stasi{\'n}ska}, G. 2005, \aap, 434, 507

\bibitem[{{Trampedach} {et~al.}(2014){Trampedach}, {Stein},
  {Christensen-Dalsgaard}, {Nordlund}, \& {Asplund}}]{2014MNRAS.445.4366T}
{Trampedach}, R., {Stein}, R.~F., {Christensen-Dalsgaard}, J., {Nordlund},
  {\AA}., \& {Asplund}, M. 2014, \mnras, 445, 4366

\bibitem[{{U} {et~al.}(2009){U}, {Urbaneja}, {Kudritzki}, {Jacobs}, {Bresolin},
  \& {Przybilla}}]{2009ApJ...704.1120U}
{U}, V., {Urbaneja}, M.~A., {Kudritzki}, R., {Jacobs}, B.~A., {Bresolin}, F.,
  \& {Przybilla}, N. 2009, \apj, 704, 1120

\bibitem[{Urbaneja {et~al.}(2005)Urbaneja, Herrero, Kudritzki, Najarro, Smartt,
  Puls, Lennon, \& Corral}]{0004-637X-635-1-311}
Urbaneja, M.~A., Herrero, A., Kudritzki, R.-P., Najarro, F., Smartt, S.~J.,
  Puls, J., Lennon, D.~J., \& Corral, L.~J. 2005, The Astrophysical Journal,
  635, 311

\bibitem[{{Urbaneja} {et~al.}(2008){Urbaneja}, {Kudritzki}, {Bresolin},
  {Przybilla}, {Gieren}, \& {Pietrzy{\'n}ski}}]{2008ApJ...684..118U}
{Urbaneja}, M.~A., {Kudritzki}, R., {Bresolin}, F., {Przybilla}, N., {Gieren},
  W., \& {Pietrzy{\'n}ski}, G. 2008, \apj, 684, 118

\bibitem[{{Verhoelst} {et~al.}(2009){Verhoelst}, {van der Zypen}, {Hony},
  {Decin}, {Cami}, \& {Eriksson}}]{2009A&A...498..127V}
{Verhoelst}, T., {van der Zypen}, N., {Hony}, S., {Decin}, L., {Cami}, J., \&
  {Eriksson}, K. 2009, \aap, 498, 127

\bibitem[{{Vilchez} {et~al.}(1988){Vilchez}, {Pagel}, {Diaz}, {Terlevich}, \&
  {Edmunds}}]{1988MNRAS.235..633V}
{Vilchez}, J.~M., {Pagel}, B.~E.~J., {Diaz}, A.~I., {Terlevich}, E., \&
  {Edmunds}, M.~G. 1988, \mnras, 235, 633

\bibitem[{{Wegner} \& {Muschielok}(2008)}]{2008SPIE.7019E..0TW}
{Wegner}, M., \& {Muschielok}, B. 2008, in Society of Photo-Optical
  Instrumentation Engineers (SPIE) Conference Series, Vol. 7019, Society of
  Photo-Optical Instrumentation Engineers (SPIE) Conference Series, 0

\bibitem[{{Westmeier} {et~al.}(2011){Westmeier}, {Braun}, \&
  {Koribalski}}]{2011MNRAS.410.2217W}
{Westmeier}, T., {Braun}, R., \& {Koribalski}, B.~S. 2011, \mnras, 410, 2217

\bibitem[{{Zurita} \& {Bresolin}(2012)}]{2012MNRAS.427.1463Z}
{Zurita}, A., \& {Bresolin}, F. 2012, \mnras, 427, 1463

\end{thebibliography}

\clearpage


\begin{deluxetable*}{cccccccc}
\tablewidth{0pt}
\tablecaption{}
\tablehead{
\colhead{Target}
& \colhead{Alt. Designation}
& \colhead{Right Ascension}
& \colhead{Declination}
& \colhead{R/R$_{25}$}
& \colhead{HST F814W}
& \colhead{IRAC Band 1}
& \colhead{IRAC Band 2} \\
 \colhead{}
& \colhead{}
& \colhead{HH:MM:SS}
& \colhead{DD:MM:SS}
& \colhead{}
& \colhead{mag}
& \colhead{mag}
& \colhead{mag}}
\startdata
 010  &                                   \nodata  &  00:54:52.07  &  $-$37:40:32.1  &  0.07  &      18.220 $\pm$  0.001  &    14.61 $\pm$  0.03   &    14.61 $\pm$  0.05  \\
 022  &                                   \nodata  &  00:54:53.83  &  $-$37:40:08.6  &  0.12  &      18.956 $\pm$  0.002  &    15.88 $\pm$  0.03   &    16.03 $\pm$  0.05  \\
 024  &                                   \nodata  &  00:54:53.12  &  $-$37:39:28.2  &  0.20  &      18.977 $\pm$  0.002  &    15.78 $\pm$  0.02   &    15.47 $\pm$  0.07  \\
 007  &        J00544792$-$3739105\tablenotemark{a}  &  00:54:47.90  &  $-$37:39:10.0  &  0.25  &      17.858 $\pm$  0.001  &    15.35 $\pm$  0.02   &    15.45 $\pm$  0.05  \\
 013  &                                   \nodata  &  00:54:49.58  &  $-$37:38:50.5  &  0.28  &      18.534 $\pm$  0.001  &              \nodata   &              \nodata  \\
 127  &        J00544267$-$3739172\tablenotemark{a}  &  00:54:42.64  &  $-$37:39:16.9  &  0.29  &                 \nodata   &    14.71 $\pm$  0.02   &    14.76 $\pm$  0.04  \\
 011  &                                   \nodata  &  00:54:49.91  &  $-$37:38:27.3  &  0.33  &      18.471 $\pm$  0.001  &    15.90 $\pm$  0.02   &    15.84 $\pm$  0.04  \\
 009  &               MPG2004 72\tablenotemark{b}  &  00:54:45.40  &  $-$37:38:35.2  &  0.33  &      18.190 $\pm$  0.001  &    15.66 $\pm$  0.04   &    15.70 $\pm$  0.05  \\
 014  &                                   \nodata  &  00:54:48.56  &  $-$37:38:02.0  &  0.38  &      18.617 $\pm$  0.001  &    15.46 $\pm$  0.02   &    15.29 $\pm$  0.06  \\
 006  &                                   \nodata  &  00:54:47.96  &  $-$37:37:39.2  &  0.43  &      17.782 $\pm$  0.001  &    15.67 $\pm$  0.02   &    15.72 $\pm$  0.04  \\
 101  &                                   \nodata  &  00:55:02.12  &  $-$37:37:50.3  &  0.49  &                 \nodata   &    15.28 $\pm$  0.02   &    15.30 $\pm$  0.05  \\
 102  &                                   \nodata  &  00:54:53.72  &  $-$37:36:47.5  &  0.55  &                 \nodata   &    15.22 $\pm$  0.02   &    15.30 $\pm$  0.04  \\
 133  &        J00543092$-$3737570\tablenotemark{a}  &  00:54:30.96  &  $-$37:37:57.0  &  0.56  &                 \nodata   &    15.57 $\pm$  0.02   &    15.58 $\pm$  0.05  \\
 130  &        J00542809$-$3738431\tablenotemark{a}  &  00:54:28.10  &  $-$37:38:42.8  &  0.57  &                 \nodata   &              \nodata   &              \nodata  \\
 126  &                                   \nodata  &  00:54:25.54  &  $-$37:39:54.6  &  0.59  &                 \nodata   &    15.22 $\pm$  0.02   &    15.26 $\pm$  0.02  \\
 135  &                                   \nodata  &  00:54:29.38  &  $-$37:37:55.9  &  0.59  &                 \nodata   &    15.26 $\pm$  0.02   &    15.35 $\pm$  0.06  \\
 134  &               MPG2004 31\tablenotemark{b}  &  00:54:28.25  &  $-$37:37:54.3  &  0.61  &                 \nodata   &              \nodata   &              \nodata  \\
 039  &                                   \nodata  &  00:54:25.76  &  $-$37:37:58.7  &  0.65  &      18.367 $\pm$  0.003  &    15.99 $\pm$  0.02   &    15.83 $\pm$  0.07  \\
 128  &                                   \nodata  &  00:54:22.00  &  $-$37:39:08.0  &  0.67  &                 \nodata   &    14.91 $\pm$  0.04   &    14.87 $\pm$  0.05  \\
 129  &        J00541708$-$3738551\tablenotemark{a}  &  00:54:17.06  &  $-$37:38:54.6  &  0.78  &                 \nodata   &              \nodata   &              \nodata  \\
 139  &                                   \nodata  &  00:54:26.77  &  $-$37:35:04.3  &  0.87  &      18.283 $\pm$  0.001  &              \nodata   &              \nodata  \\
 034  &               MPG2004 15\tablenotemark{b}  &  00:54:18.59  &  $-$37:35:38.4  &  0.92  &      18.624 $\pm$  0.001  &              \nodata   &              \nodata  \\
 055  &                                   \nodata  &  00:54:22.70  &  $-$37:34:40.0  &  0.95  &      19.012 $\pm$  0.001  &              \nodata   &              \nodata  \\
 132  &                                   \nodata  &  00:54:08.74  &  $-$37:38:02.9  &  0.96  &                 \nodata   &              \nodata   &              \nodata  \\
 140  &        J00541835$-$3735024\tablenotemark{a}  &  00:54:18.34  &  $-$37:35:02.0  &  0.97  &                 \nodata   &              \nodata   &              \nodata  \\
 031  &                                   \nodata  &  00:54:23.27  &  $-$37:34:20.3  &  0.97  &      18.358 $\pm$  0.001  &              \nodata   &              \nodata  \\
 037  &                                   \nodata  &  00:54:16.18  &  $-$37:34:51.6  &  1.01  &      18.907 $\pm$  0.002  &              \nodata   &              \nodata 
\enddata
\tablecomments{For objects with no HST F814W entry HST photometry was either not available or uncertain.}
\label{tbl:fitpar}
\tablenotetext{a}{\cite{2006AJ....131.1163S}}
\tablenotetext{b}{\cite{2004MNRAS.350..679M}}
\end{deluxetable*}

\begin{deluxetable*}{ccc}
\tablewidth{0pt}
\tablecaption{}
\tablehead{
\colhead{Target}
& \colhead{Mass}
& \colhead{Luminosity} \\
 \colhead{}
& \colhead{M$_\odot$}
& \colhead{log L/L$_\odot$}}
\startdata
 010  &    17.4 $\pm$ 1.0  &    4.92 $\pm$  0.08 \\
 022  &    12.3 $\pm$ 0.8  &    4.64 $\pm$  0.08 \\
 024  &    14.3 $\pm$ 0.8  &    4.64 $\pm$  0.08 \\
 007  &    18.8 $\pm$ 1.1  &    5.05 $\pm$  0.08 \\
 013  &    16.1 $\pm$ 0.9  &    4.80 $\pm$  0.08 \\
 127  &          \nodata   &            \nodata  \\
 011  &    16.3 $\pm$ 0.9  &    4.82 $\pm$  0.08 \\
 009  &    17.5 $\pm$ 1.0  &    4.93 $\pm$  0.08 \\
 014  &    15.5 $\pm$ 0.9  &    4.77 $\pm$  0.08 \\
 006  &    19.2 $\pm$ 1.2  &    5.08 $\pm$  0.08 \\
 101  &          \nodata   &            \nodata  \\
 102  &          \nodata   &            \nodata  \\
 133  &          \nodata   &            \nodata  \\
 130  &          \nodata   &            \nodata  \\
 126  &          \nodata   &            \nodata  \\
 135  &          \nodata   &            \nodata  \\
 134  &          \nodata   &            \nodata  \\
 039  &    16.6 $\pm$ 0.9  &    4.86 $\pm$  0.08 \\
 128  &          \nodata   &            \nodata  \\
 129  &          \nodata   &            \nodata  \\
 139  &    16.9 $\pm$ 1.0  &    4.89 $\pm$  0.08 \\
 034  &    15.8 $\pm$ 0.8  &    4.77 $\pm$  0.08 \\
 055  &    14.4 $\pm$ 0.8  &    4.62 $\pm$  0.08 \\
 132  &          \nodata   &            \nodata  \\
 140  &          \nodata   &            \nodata  \\
 031  &    16.7 $\pm$ 1.0  &    4.86 $\pm$  0.08 \\
 037  &    14.8 $\pm$ 0.8  &    4.66 $\pm$  0.08
\enddata
\tablecomments{}
\label{tbl:fitpar}
\end{deluxetable*}

\end{document}